\documentclass[english,aps,superscriptaddress,showkeys,showpacs,prepri,twocolumn]{revtex4}
\usepackage[T1]{fontenc}
\pdfoutput=1
\usepackage[letterpaper]{geometry}
\usepackage{units}
\usepackage{bbding}
\usepackage{amsmath}
\usepackage{amssymb}
\usepackage{graphicx}
\usepackage{esint}
\usepackage{sidecap}
\usepackage[colorinlistoftodos,prependcaption,textsize=tiny]{todonotes}

\usepackage[utf8]{inputenc}
\usepackage{lmodern} 

\makeatletter

\@ifundefined{textcolor}{}
{%
 \definecolor{BLACK}{gray}{0}
 \definecolor{WHITE}{gray}{1}
 \definecolor{RED}{rgb}{1,0,0}
 \definecolor{GREEN}{rgb}{0,1,0}
 \definecolor{BLUE}{rgb}{0,0,1}
 \definecolor{CYAN}{cmyk}{1,0,0,0}
 \definecolor{MAGENTA}{cmyk}{0,1,0,0}
 \definecolor{YELLOW}{cmyk}{0,0,1,0}
}

\usepackage{doi}
\usepackage{hyperref}

\makeatother

\usepackage{babel}
\begin{document}

\title{NIMROD Modeling of Quiescent H-mode: Reconstruction Considerations and
Saturation Mechanism}

\author{J. R. King}
\affiliation{Tech-X Corporation, 5621 Arapahoe Ave. Boulder, CO 80303, USA}

\author{K. H. Burrell}
\affiliation{General Atomics, PO Box 85608, San Diego, CA 92186–5608, USA}

\author{A. M. Garofalo}
\affiliation{General Atomics, PO Box 85608, San Diego, CA 92186–5608, USA}

\author{R. J. Groebner}
\affiliation{General Atomics, PO Box 85608, San Diego, CA 92186–5608, USA}

\author{S. E. Kruger}
\affiliation{Tech-X Corporation, 5621 Arapahoe Ave. Boulder, CO 80303, USA}

\author{A. Y. Pankin}
\affiliation{Tech-X Corporation, 5621 Arapahoe Ave. Boulder, CO 80303, USA}

\author{P. B. Snyder}
\affiliation{General Atomics, PO Box 85608, San Diego, CA 92186–5608, USA}

\date{draft \today}
\begin{abstract}
The extended-MHD NIMROD code [C.R.~Sovinec and J.R.~King, J.~Comput.~Phys.~{\bf
229}, 5803 (2010)] models broadband-MHD activity from a reconstruction of a
quiescent H-mode shot on the DIII-D tokamak [J. L. Luxon, Nucl. Fusion 42, 614
(2002)]. Computations with the reconstructed toroidal and
poloidal ion flows exhibit low-$n_\phi$ perturbations ($n_\phi\simeq1-5$) that
grow and saturate into a turbulent-like MHD state. The workflow used to project
the reconstructed state onto the NIMROD basis functions re-solves the
Grad-Shafranov equation and extrapolates profiles to include scrape-off-layer
currents. Evaluation of the transport from the turbulent-like MHD state leads
to a relaxation of the density and temperature profiles. 
Published version: Nucl. Fusion 57 022002 (2017) [\url{http://dx.doi.org/10.1088/0029-5515/57/2/022002}]
\end{abstract}


\maketitle



  \newcommand{\vect}[1]{ \mathbf{#1}}
  \newcommand{\defn}{ \equiv}

  \newcommand{\lp}{\left(}
  \newcommand{\rp}{\right)}
  \newcommand{\lb}{\left[}
  \newcommand{\rb}{\right]}
  \newcommand{\la}{\left<}
  \newcommand{\ra}{\right>}

  \newcommand{\vf}{ \vect{f}}

  \newcommand{\vx}{\vect{x}}
  \newcommand{\vq}{\vect{q}}
  \newcommand{\vB}{\vect{B}}
  \newcommand{\vJ}{\vect{J}}
  \newcommand{\vA}{\vect{A}}
  \newcommand{\vE}{\vect{E}}
  \newcommand{\vV}{\vect{V}}
  \newcommand{\vF}{ \vect{F} }	
  \newcommand{\vU}{ \vect{U} }	
  \newcommand{\ddp}{\grad \cdot \Pi}
  \newcommand{\specheat}{\gamma_h}

  \newcommand{\grad}{\vect{\nabla}}
  \newcommand{\curl}[1]{\grad \times #1 }
  \newcommand{\dive}[1]{\grad \cdot #1 }
  \newcommand{\vdg}{\left(\vV \cdot \grad \right)}
  \newcommand{\bdg}{\left(\vB \cdot \grad \right)}
  \newcommand{\divV}{\grad \cdot \vV_1}
  \newcommand{\divVp}{\left( \grad \cdot \vV \right)}

  \newcommand{\dt}[1]{\frac{\partial #1}{\partial t}}
  \newcommand{\Dt}[1]{\frac{d #1}{dt}}
  \newcommand{\dpsi}[1]{\frac{\partial #1}{\partial \psi}}
  \newcommand{\dpsisq}[1]{\frac{\partial^2 #1}{\partial \psi^2}}
  
  \newcommand{\jac}{{\mathcal{J}}}
  \newcommand{\jaci}{{\mathcal{J}}^{-1}}
  \newcommand{\Pp}{ P^\prime }				
  \newcommand{\Vp}{V^\prime}
  \newcommand{\Vpp}{V^{\prime\prime}}
  \newcommand{\Vpo}{ \frac{V^\prime}{4 \pi^2}}
  \newcommand{\norm}{ P^\prime }		
  \newcommand{\RR}{ \psi }			
  \newcommand{\vR}{ \grad \RR }		
  \newcommand{\C}{ C }				
  \newcommand{\vC}{ \vect{\C} }		
  \newcommand{\vK}{ \vect{K} }			
  \newcommand{\vRsq}{ \mid \grad \RR \mid^2 }
  \newcommand{\vCsq}{ \C^2 }
  \newcommand{\vKsq}{ K^2 }
  \newcommand{\vBsq}{ B^2 }
  \newcommand{\vrr}{\frac{ \vR}{\vRsq} }
  \newcommand{\vbb}{\frac{ \vB}{B^2} }
  \newcommand{\vcc}{\frac{ \vC}{\vCsq} }
  \newcommand{\vjj}{\frac{ \vJ}{J^2} }
  \newcommand{\vkk}{\frac{ \vK}{\vKsq} }

  \newcommand{\R}{ \psi }
  \newcommand{\T}{ \Theta }
  \newcommand{\Z}{ \zeta }
  \newcommand{\A}{ \alpha }
  \newcommand{\U}{ u }
  \newcommand{\ve}{ \vect{e} }
  \newcommand{\vur}{ \vect{e}^\rho }
  \newcommand{\vut}{ \vect{e}^\Theta }
  \newcommand{\vuz}{ \vect{e}^\zeta }
  \newcommand{\vlr}{ \vect{e}_\rho }
  \newcommand{\vlt}{ \vect{e}_\Theta }
  \newcommand{\vlz}{ \vect{e}_\zeta }
  \newcommand{\gr}{ \grad \R }
  \newcommand{\gt}{ \grad \Theta }
  \newcommand{\gz}{ \grad \zeta }
  \newcommand{\ga}{ \grad \alpha }
  \newcommand{\gu}{ \grad \U }
  \newcommand{\dr}[1]{ \frac{\partial #1}{\partial \R} }
  \newcommand{\dT}[1]{\frac{\partial #1}{\partial \Theta}}
  \newcommand{\dz}[1]{\frac{\partial #1}{\partial \zeta}}
  \newcommand{\dU}[1]{\frac{\partial #1}{\partial \U}}
  \newcommand{\drs}[1]{ \frac{\partial^2 #1}{\partial \R^2} }
  \newcommand{\dTs}[1]{\frac{\partial^2 #1}{\partial \Theta^2}}
  \newcommand{\drt}[1]{\frac{\partial^2 #1}{\partial \R \partial \Theta}}
  \newcommand{\dzs}[1]{\frac{\partial^2 #1}{\partial \zeta^2}}
  \newcommand{\grr}{ g^{\R \R} }
  \newcommand{\grt}{ g^{\R \Theta} }
  \newcommand{\grz}{ g^{\R \zeta} }
  \newcommand{\gtz}{ g^{\Theta \zeta} }
  \newcommand{\gtt}{ g^{\Theta \Theta} }
  \newcommand{\gzz}{ g^{\zeta \zeta} } 
  \newcommand{\ri}{ \frac{1}{R^2} }
  \newcommand{\fr}{ \lp \R \rp}
  \newcommand{\frt}{ \lp \R, \T \rp}
  \newcommand{\frtz}{ \lp \R,\T,\Z \rp}

  \newcommand{\fluxav}[1]{\la #1 \ra}
  \newcommand{\thetaav}[1]{\la #1 \ra_\T}

\newcommand{\cramplist}{
        \setlength{\itemsep}{0in}
        \setlength{\partopsep}{0in}
        \setlength{\topsep}{0in}}
\newcommand{\cramp}{\setlength{\parskip}{.5\parskip}}
\newcommand{\zapspace}{\topsep=0pt\partopsep=0pt\itemsep=0pt\parskip=0pt}

\section{Introduction}
\label{sec:introduction}


It is desirable to have an ITER \cite{ITER} H-mode regime that is quiescent to
edge-localized modes (ELMs) \cite{connor98,leonard06}. ELMs deposit large,
localized and impulsive heat loads that can damage the divertor. A quiescent
regime with edge harmonic oscillations (EHO) or broadband MHD activity is
observed in some DIII-D
\cite{burrell01,burrell05,burrell09,garofalo11,burrell12,burrell13,solomon14,garofalo15},
JT-60U \cite{Sakamoto04,oyama05}, JET \cite{solano10} and ASDEX-U \cite{suttrop05}, discharge
scenarios. These ELM-free discharges have the pedestal-plasma confinement
necessary for burning-plasma operation in ITER\cite{garofalo15}. The mode
activity associated with the EHO or broadband MHD on DIII-D is characterized by
small toroidal-mode numbers ($n_\phi\simeq1-5$) and is thus suitable for
simulation with global MHD codes.  Measurements from beam-emission
spectroscopy, electron-cyclotron emission, and magnetic probe diagnostics show
highly coherent density, temperature and magnetic oscillations associated with
EHO.  The particle and impurity \cite{grierson15} transport is enhanced during
QH-mode, leading to essentially steady-state profiles in the pedestal region.

Relative to QH-mode operation with EHO, operation with broadband MHD tends to
occur at higher densities and lower rotation and thus may be more relevant to
potential ITER discharge scenarios.  While there are computational
investigations of the discharges with EHO \cite{liu15,battaglia14}, there is
less computational analysis of discharges with broadband MHD. In this paper, we
investigate the broadband-MHD state with nonlinear NIMROD
\cite{Sovinec04,Sovinec10} simulations initialized from a reconstruction of a
DIII-D QH-mode discharge with broadband MHD. These simulations include the
reconstructed flow and saturate into a turbulent-like state.

This paper is organized as follows.  As an initial-value computation, our
simulations require accurate and smooth initial conditions to avoid 
spurious instabilities. 
Section \ref{sec:reconstruct} describes how the reconstructed
fields are imported into the NIMROD spatial discretization. One of the novel
methods in our approach to this edge modeling is to extrapolate the pressure
and density profiles through the scrape-off layer (SOL) to maintain first-order
continuity and thus a continuous current profile when solving the
Grad-Shafranov equation.  The section concludes by discussing our assumptions
where the reconstructed fields are in steady state, and how our modeling
simulates the dynamics of perturbations about this steady state. In
Sec.~\ref{sec:nonlinear}, the model equations and the dynamics of our nonlinear
MHD simulations that saturate into a turbulent-like state are considered. The
magnetic stochasticity and transport induced from these perturbations are
analyzed in Sec.~\ref{sec:transport}.  Finally, we conclude with a discussion
of the implications and limitations of our present modeling and mention future
directions for this work.
 

\section{Extended equilibrium reconstruction for NIMROD}
\label{sec:reconstruct}

\begin{figure}
  \centering
  \includegraphics[width=8cm]{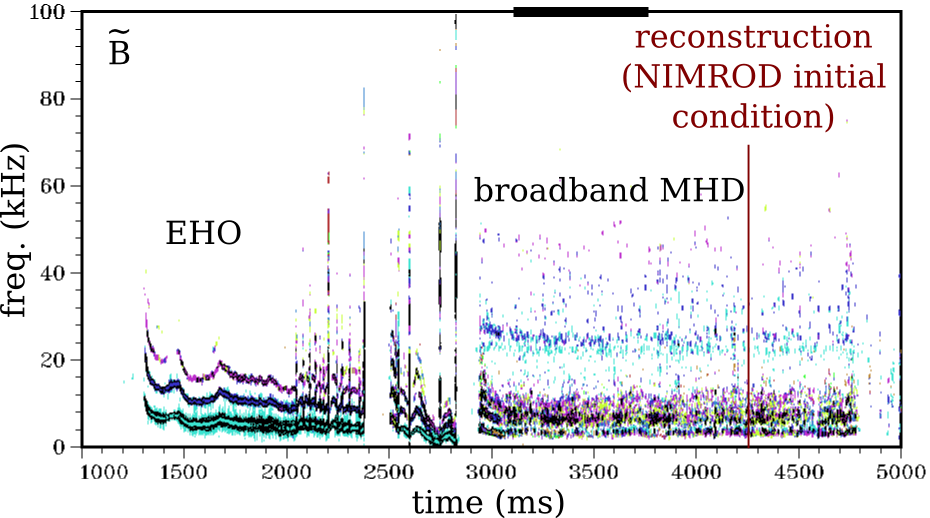}
  \vspace{-4mm}
  \caption{[Color online]
  Cross-power spectrum plot of the magnetic fluctuation probe measurements
  from DIII-D shot 145098. An initial phase contains coherent EHO fluctuations
  followed by a phase with broadband-MHD activity. The NIMROD simulation
  is initialized from a reconstruction during the latter phase at 4250 ms.
  }
  \label{fig:newspec}
\end{figure}

The cross-power spectrum from magnetic fluctuation probe measurements on DIII-D
during shot 145098 is shown in Fig.~\ref{fig:newspec}.  This shot is part of a
campaign to create a QH-mode discharge with ITER-relevant parameters and shape,
the ultimate results of which are summarized in Ref.~\cite{garofalo15}.  As the
torque is ramped down the mode activity transitions from EHO to broadband MHD.
An EFIT \cite{lao85,lao05} reconstruction, constrained by magnetic-probe,
motional-stark-effect, Thomson-scattering, and charge-exchange-recombination
(CER) measurements, is used to specify the initial condition in the NIMROD code.
High quality equilibria are essential for extended-MHD modeling with
initial-value codes such as NIMROD.  Typically the spatial resolution
requirements for extended-MHD modeling, which must resolve singular-layer
physics and highly anisotropic diffusion, are more stringent than the
resolution of equilibrium reconstructions from experimental discharges. To
circumvent mapping errors, we re-solve the Grad-Shafranov equation with
open-flux regions using the NIMEQ \cite{Howell14} solver to generate a new
equilibrium while using the mapped results for both an initial guess and to
specify the boundary condition. 

\begin{SCfigure}
  \includegraphics[width=4cm]{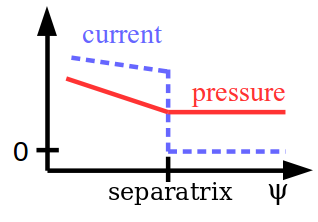}
  \caption{[Color online]
  Sketch that shows a discontinuous first derivative in pressure
  causes a discontinuous current profile when solving the Grad-Shafranov 
  equation.}
  \label{fig:jp-sketch}
\end{SCfigure}

Additionally, reconstructions commonly assume that the region outside the last
closed flux surface (LCFS) is current free. The pressure, temperature and
density profiles are specified only up to the LCFS and are assumed to be
constant outside the LCFS as illustrated in Fig.~\ref{fig:jp-sketch}.  For
discharges with large pedestal current, as is commonly found during QH-mode,
this can lead to a large discontinuity in the current density at the LCFS that
is problematic for MHD modeling. During our re-solve of the Grad-Shafranov
equation, we relax the current-free assumption outside the LCFS and include
temperature and density profiles with non-zero gradients which generate
associated small currents in the scrape-off layer (SOL) that cause the overall
current profile to be continuous. 
Modified-bump-function fits are used to smoothly extrapolate the pressure,
electron temperature and particle density in the SOL region. Derivatives of all
orders vanish for this functional fit at the edge of the SOL region and thus
the resulting current profile smoothly decays to zero. For this case, the
pressure drops from $922$ to $581\;Pa$, the electron temperature drops from
$186$ to $30\;eV$ and the density drops from $4.6\times10^{18}$ to
$2.5\times10^{18}\;m^{-3}$ in the SOL region. The half width of the electron
pressure profile is roughly $2.5\;mm$ at the outboard mid-plane and $2\;cm$ at
the divertor plate. This results in a SOL width that is slightly smaller than
the measured half width of the heat-flux during the later half of the inter-ELM
period of DIII-D ELMy H-mode discharges in Ref.~\cite{eich13}. These profiles
fits, made with a focus on the resulting current and resistivity profiles,
result in an ion-temperature profile that remains above $1\;keV$ throughout the
domain. This inconsistency will be removed in future modeling. The new solution
is an equilibrium that closely resembles the original reconstruction with the
exception of the open-flux currents and additional quantification of these
methods will be described in a future manuscript \cite{kingSOL}. 
This regenerated equilibrium is consistent with the
core profiles that are measured by the high quality diagnostics on DIII-D.

\begin{figure}
  \centering
  \includegraphics[width=7cm]{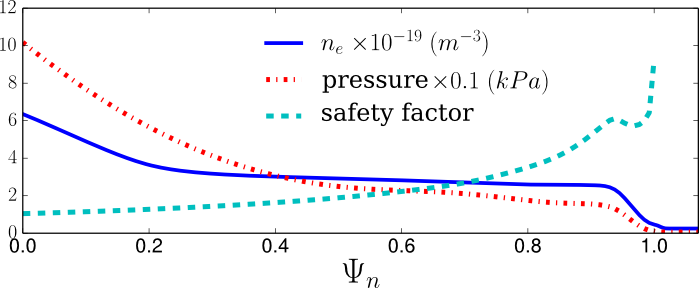}
  \vspace{-4mm}
  \caption{[Color online]
  Density, pressure and safety-factor profiles as a function of 
  normalized flux. SOL profiles for density and pressure are also
  shown for $\psi_n>1$.}
  \label{fig:npq}
\end{figure}

\begin{figure}
  \centering
  \includegraphics[width=7cm]{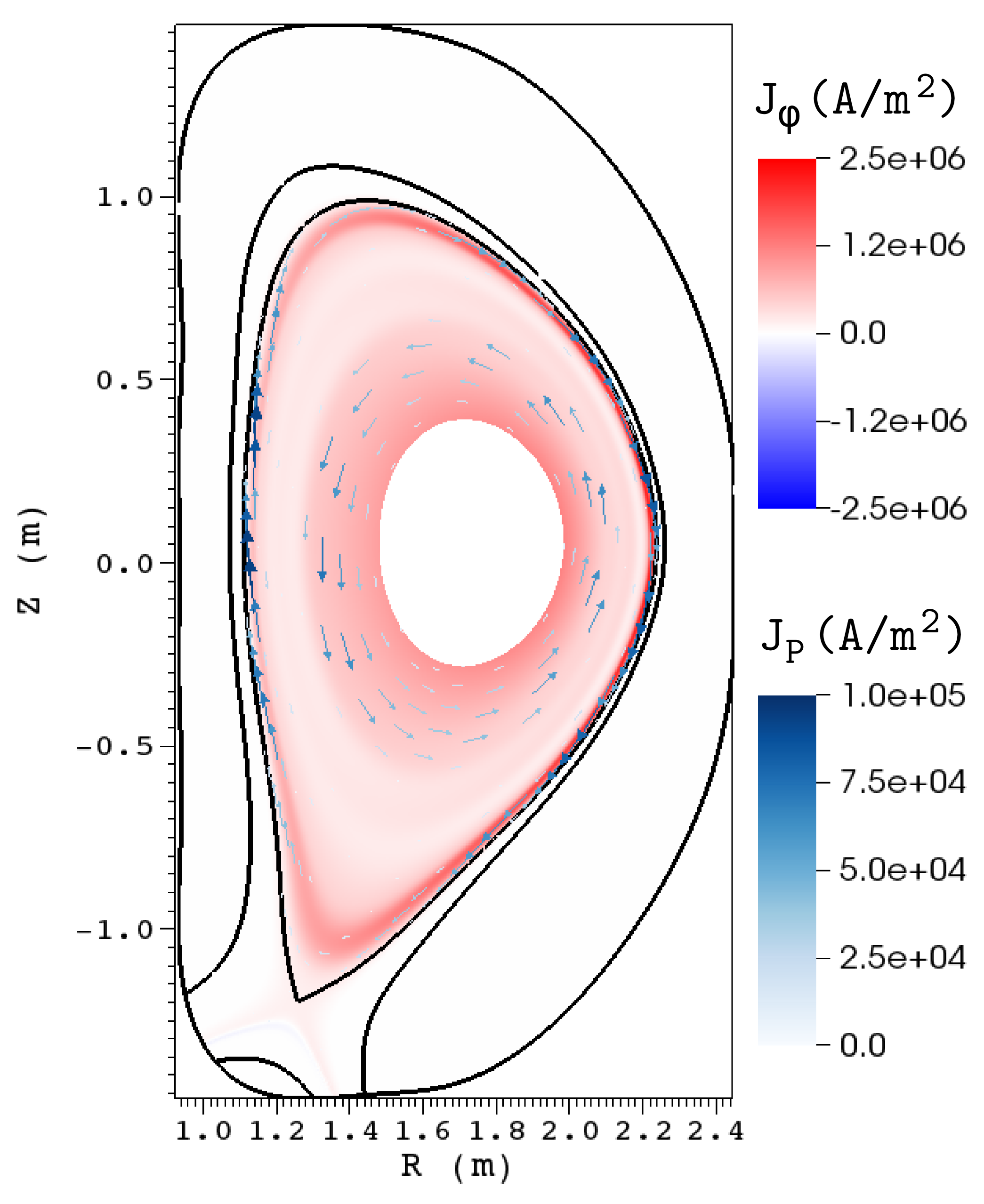}
  \vspace{-4mm}
  \caption{[Color online]
  Initial current density (contour colors for toroidal current and arrow
  vectors for poloidal current) from the reconstructed state with an extrapolated SOL
  region. The SOL region contains small, but non-zero currents that
  terminate poloidally on the divertor. }
  \label{fig:current}
\end{figure}

The density, pressure and safety-factor profiles as a function of normalized
flux are shown in Fig.~\ref{fig:npq}, where the SOL profiles for density and
pressure are included where $\psi_n>1$. The current-profile that results
from our re-solve of the Grad-Shafranov equation is plotted in Fig.~\ref{fig:current}.
The SOL region contains small, but non-zero, currents that terminate
poloidally on the divertor.

\begin{figure}
  \centering
  \includegraphics[width=7cm]{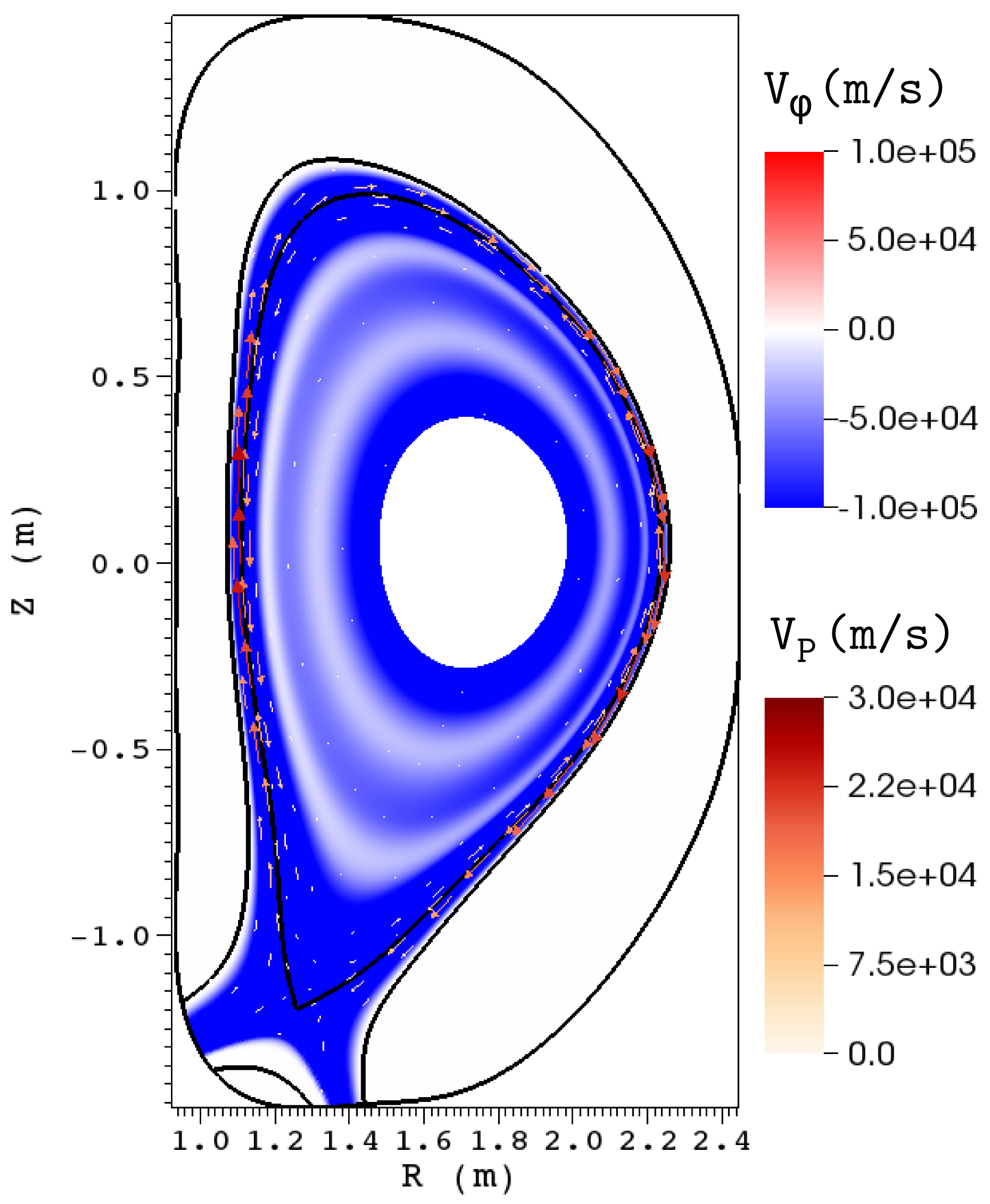}
  \vspace{-4mm}
  \caption{[Color online]
  The reconstructed toroidal (contour colors) and poloidal (arrow vector) flow with
  extrapolated flows in the SOL region. The poloidal flow flips sign just inside
  the LCFS with flows closer to the core proceeding counter-clockwise and flows
  near and outside the LCFS proceeding clockwise. }
  \label{fig:flows}
\end{figure}

Importantly, our initial conditions include the full reconstructed
toroidal and poloidal flows as shown in Fig.~\ref{fig:flows}.
Experimentally, these flows are critical to the observation of QH-mode where,
in particular, large $\mathbf{E}\times\mathbf{B}$ flow shear is highly correlated with quiescent
operation \cite{garofalo11}. 
Like the thermodynamic profiles, the flow profiles are specified up to and
are non-zero at the LCFS. Thus we extrapolate these profiles to zero within the
SOL.

%

Typically only MHD-force balance (a Grad-Shafranov solution) is strictly
enforced for the steady state.  In practice, perturbations about a
time-independent equilibrium are evolved, and that the time-independent
equilibrium need not be a time-independent solution of the source-free
resistive MHD equations \cite{Sovinec04,charlton86}. This
effectively assumes the presence of implicit (in the sense that they are
calculable but not calculated) sources, fluxes and sinks.  With these
assumptions, if the code is run on a MHD-stable case with $n_\phi>0$
perturbations, the modification to the $n_\phi=0$ fields is insignificant.
Alternatively, when the case is MHD-unstable, the initial $n_\phi=0$ fields are
self-consistently modified by the presence of the unstable modes.

The NIMROD
code has the capability to compute the extended-MHD evolution of the
reconstructed fields. However, it is well-known that physical mechanisms
outside the scope of our modeling equations mediate tokamak transport such as
neoclassical bootstrap current, toroidal viscosity, and poloidal flow damping,
neutral beam and RF drives, kinetic turbulence, and coupling to the scrape-off layer
(SOL), neutrals, impurities and the material boundary. Including these effects
requires explicit calculation of the sources, fluxes and sinks. These
transport-type calculations are possible and are becoming practical (e.g.
\cite{Jenkins12,Jenkins15,held15}), but this sort of integrated modeling
remains in the future. Thus in this work we assume that the initial
reconstructed fields are steady state and our goal is to model the evolution of
the 3D perturbations around this state.

\section{Nonlinear evolution}
\label{sec:nonlinear}

\begin{figure}
  \includegraphics[width=8cm,trim={0.5cm 2.5cm 2cm 2cm},clip]{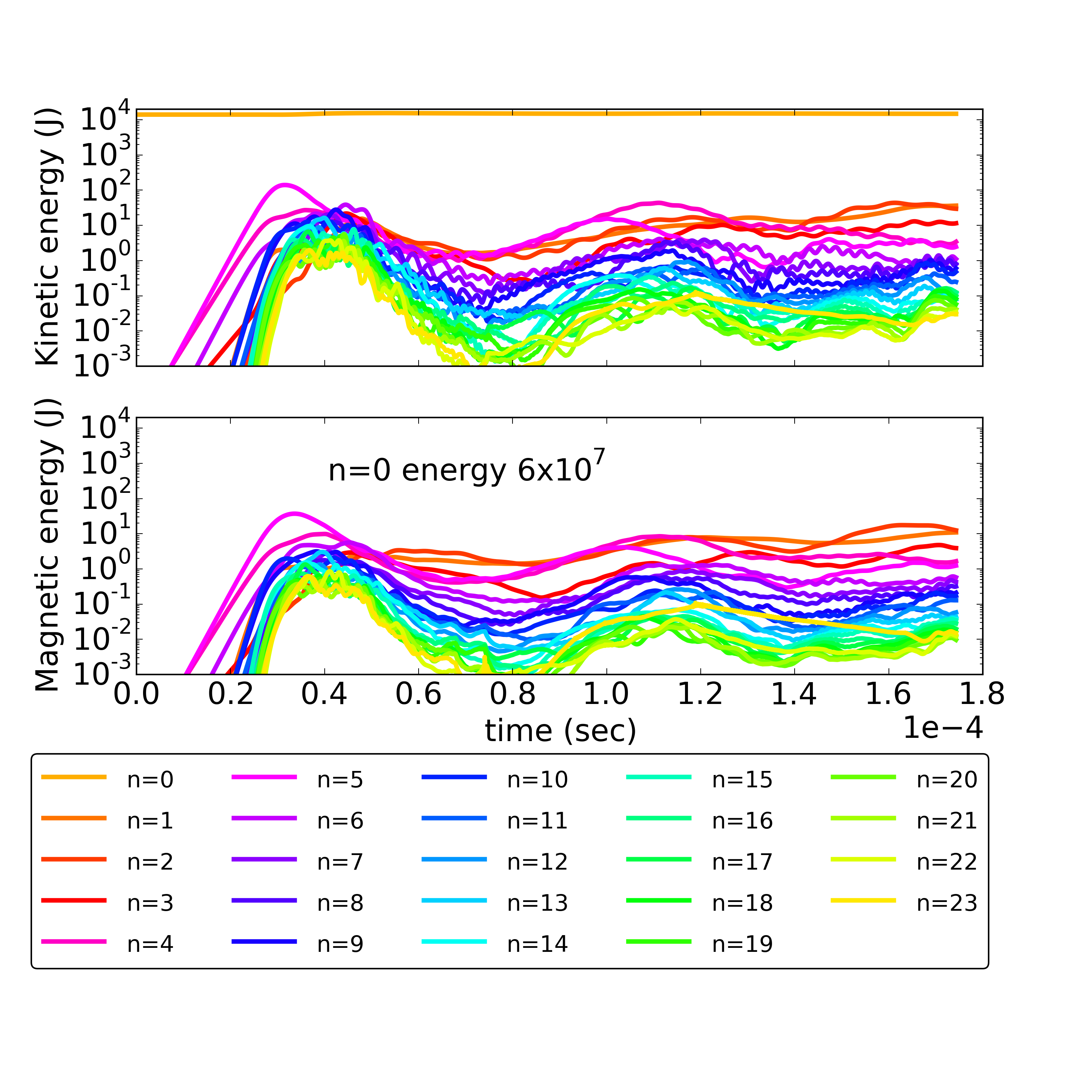}
  \vspace{-6mm}
  \caption{[Color online]
  Simulated kinetic and magnetic energy decomposed by toroidal mode number ($n_\phi$).}
  \label{fig:energy}
\end{figure}

\begin{figure*}
  \centering
  \includegraphics[width=16cm]{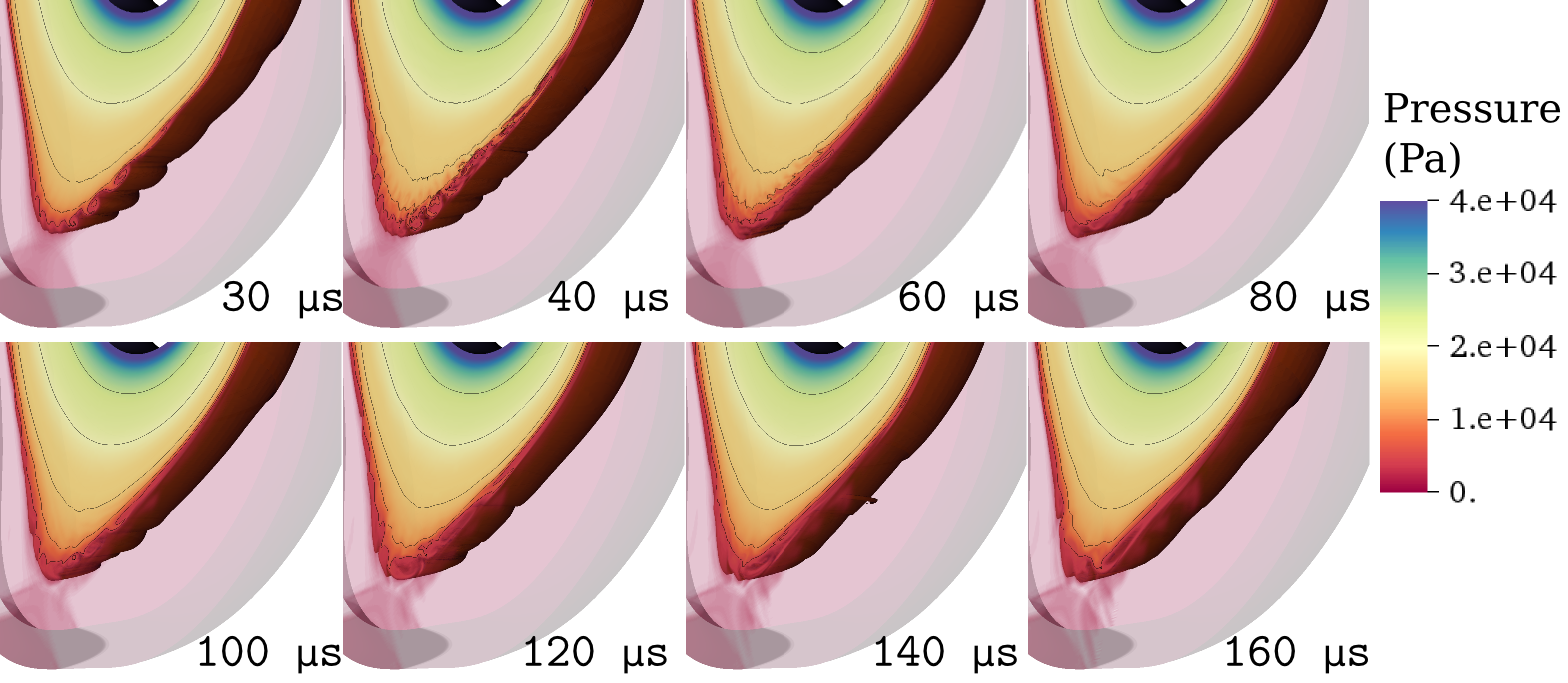}
  \vspace{-4mm}
  \caption{[Color online]
  A slice of the 3D domain that shows the evolution of the pressure contours at
eight different times. During the initial stages (t=$30$ and $40\;\mu s$), eddies
of hot, high density plasma are ejected from the pedestal and are advected
poloidally in the counter-clockwise direction. Later in time these eddies are
sheared apart and the dynamics become a more turbulent-like state with smoke-like
off gassing which precesses poloidally counter-clockwise. }
  \label{fig:pres3D}
\end{figure*}

The nonlinear simulation uses a single-fluid, single-temperature (assuming fast
equilibration in the perturbations) MHD model with a temperature-dependent
resistivity profile where the Lundquist number, $S$, in the core is
$1.1\times10^6$. Here $S=\tau_R/\tau_A$, where $\tau_A$ is the Alfvén time
($\tau_A=R_o/v_A$), $v_A$ is the Alfvén velocity ($B/\sqrt{m_i n_i \mu_0}$),
$\tau_R$ is the resistive diffusion time ($\tau_R=R_o^2 \mu_0/\eta$),
$R_o=1.748 m$ is the radius of the magnetic axis, $\eta$ is the electrical
resistivity, $\mu_0$ is the permeability of free space, $m_i$ is the ion mass,
and $n_i$ is the ion density.  This choice of resistivity is enhanced by a
factor of 100 relative to the Spitzer value for computational practicality.
The model includes large parallel and small perpendicular diffusivities in 
the momentum and energy equations. The parallel-momentum-stress contribution is 
\begin{equation}
\mathbf{\Pi}_{\parallel i}=
  m_i n_i \nu_{\parallel i}
  \left(\hat{\mathbf{b}}
  \hat{\mathbf{b}}-\frac{1}{3}\mathbf{I}\right)\left(
  3\hat{\mathbf{b}}\cdot\nabla\mathbf{v}_{i}\cdot
  \hat{\mathbf{b}}-\nabla\cdot\mathbf{v}_{i}\right)\;,
\end{equation} where $\mathbf{v}_i$ is the ion velocity,
$\hat{\mathbf{b}}=\mathbf{B}/|B|$, $\mathbf{I}$ is the identity tensor, and
$\nu_{\parallel i}=10^5\;m^2/s$. The parallel-heat-flux contribution is
\begin{equation}
\mathbf{q}_{\parallel}=
  -n_i \chi_{\parallel}
  \hat{\mathbf{b}}\hat{\mathbf{b}}\cdot\nabla T\;,
\end{equation}
where $T$ is the temperature and $\chi_{\parallel}=10^8\;m^2/s$.
The small perpendicular diffusivites are modeled as isotropic particle,
momentum and thermal diffusivities with a magnitude of $1\;m^2/s$.


The 3D nonlinear simulation is performed with a $60\times128$ high-order
(biquartic) finite element mesh packed around the pedestal region to resolve
the poloidal plane and $24$ Fourier modes in the toroidal direction. The
simulation is initialized from a linear computation of modes with a restricted
toroidal mode number range ($n_\phi=1-8$). The mode energies at $t=0s$ are
small, the largest energy is contained within the $n_\phi=4$ mode which has a
spectral kinetic energy content of $4.2\times10^{-5}\;J$ and a spectral
magnetic energy content of $4.4\times10^{-6}\;J$.

The boundary conditions, on both the inner annulus and outer wall, are no-slip
for the velocity, Dirichlet for the density and temperature and a perfectly
conducting wall boundary condition for the magnetic field.  Linear computations
show that the mode growth rates are unaffected by presence of the inner
boundary, however there is an important effect in nonlinear computations.  The
Dirichlet condition on density and temperature provides an unrestricted
particle and energy source in the core to maintain the profiles at the inner
boundary in the presence of fluctuation-induced transport. With respect to the
outer boundary, a sheath boundary condition is not applied at the divertor and
consideration of an improved divertor boundary condition is a direction for future
research.


The energy evolution from a nonlinear NIMROD simulation, decomposed by toroidal
mode number, of DIII-D QH-mode shot 145098 at $4250\;ms$ with broadband MHD
activity is shown in Fig~\ref{fig:energy}.   The simulations are initially
dominated by a $n_\phi=5$ perturbation that saturates at around $30\;\mu s$.
After this time a saturated turbulent-like state develops and the $n_\phi=1$
and $2$ modes become dominant through an inverse cascade.  Each toroidal mode
in the range of $n_\phi=1-5$ is dominant at a different time and continued
interplay between modes is observed as the simulation progresses, particularly in 
the kinetic energy spectrum.
Figure~\ref{fig:pres3D} plots the lower half of a poloidal cut of the 3D
pressure contours at eight different time slices. The first time slice ($30\;\mu
s$) shows a coherent
structure associated with the dominant $n_\phi=5$ mode. By $40\;\mu s$, this
structure becomes sheared apart leading to a turbulent-like state at later
times.  Higher-time resolution plots show the perturbations are advected 
in the counter-clockwise direction consistent with the direction
of the ion poloidal flow inside the LCFS with a smoke-like off-gassing behavior.

\section{Transport induced by the MHD perturbations}
\label{sec:transport}

\begin{figure}
  \centering
  \includegraphics[width=8cm]{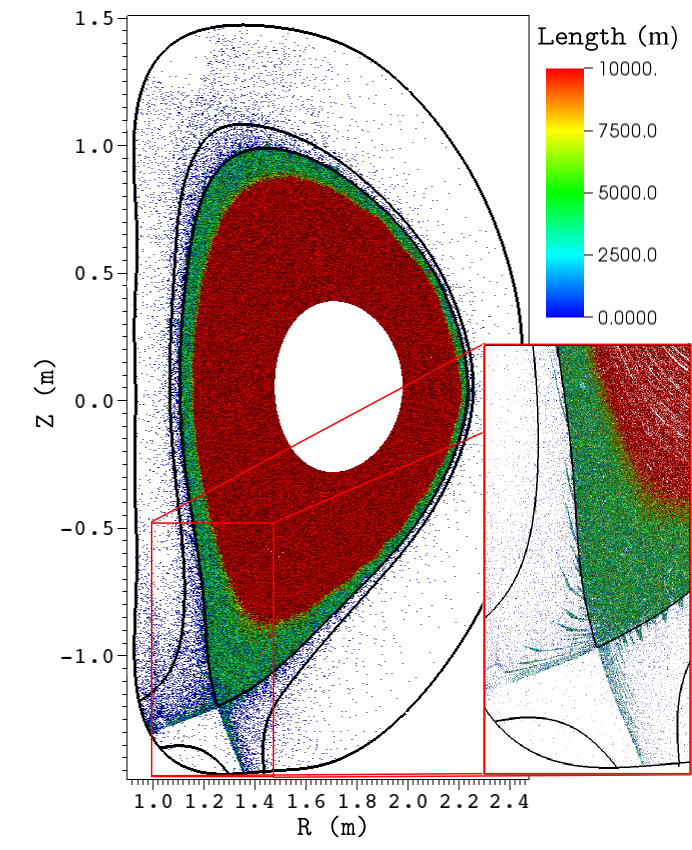}
  \vspace{-4mm}
  \caption{[Color online]
  Poincaré plot of magnetic fieldlines at t=$40\;\mu s$ . The pedestal region
  becomes stochastic where the original LCFS and SOL-region bounding contours are
  shown for reference. Inset shows homoclinic tangle structure near the divertor
  x-point.  }
  \label{fig:fieldline}
\end{figure}

In the presence of these electro-magnetic MHD perturbations, the pedestal
region becomes stochastic as shown in Fig.~\ref{fig:fieldline} at t=$40\;\mu s$
by a magnetic field-line Poincaré (or puncture) plot. Field-lines in this
figure are followed for $10^4\;m$ or until they hit the wall and then are color
coded by their total length.  The region from near the top of the pedestal out
to the LCFS becomes stochastic. As the inset figure shows, a homoclinic tangle
structure \cite{evans02,roeder03,evans05} develops near the divertor x-point.
Given the large parallel thermal conductivity and stochastic magnetic fields,
one expects significant energy transport within the pedestal region to result.
However, as shown and discussed next, this is not the case and the energy
transport is relatively small when compared with the particle transport.

\begin{figure}
  \centering
  \includegraphics[width=7cm]{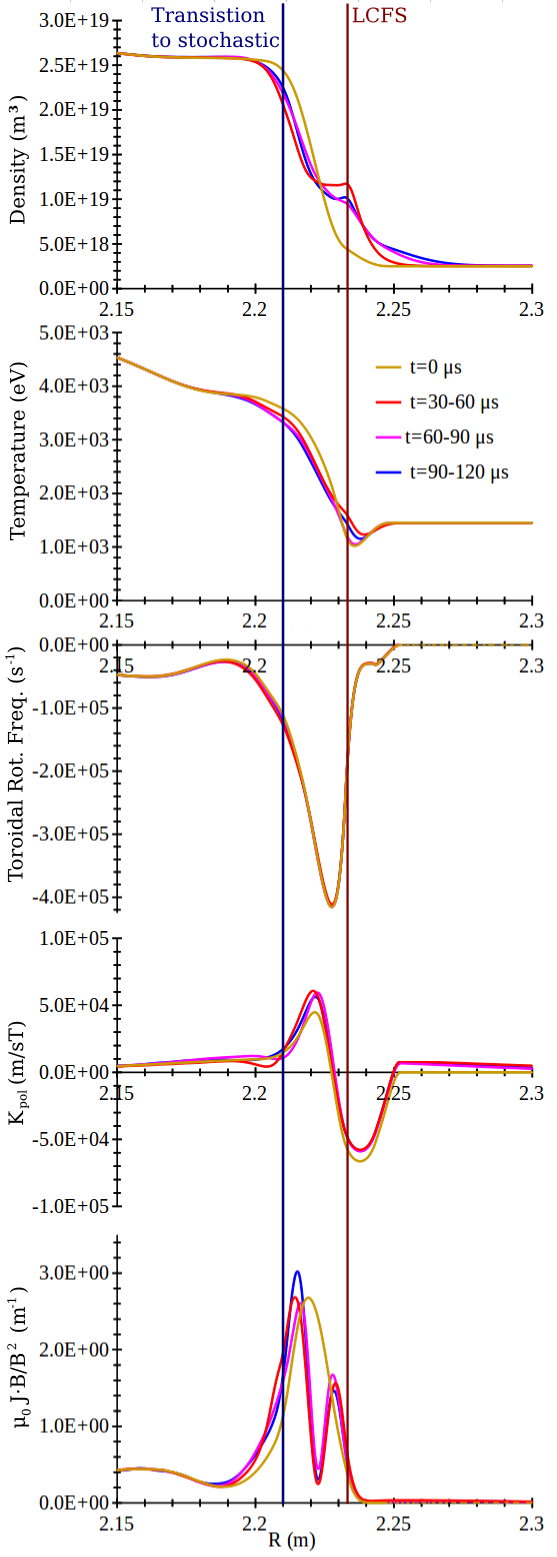}
  \vspace{-4mm}
  \caption{[Color online]
  Toroidal average of the density, temperature, toroidal- and
  poloidal-rotation, and current-density profiles on the outboard midplane for
  the initial condition and average values during three time windows. }
  \label{fig:profs}
\end{figure}

Figure~\ref{fig:profs} shows the toroidal average of the density, temperature,
toroidal- and poloidal-rotation, and current-density profiles on the outboard
midplane for the initial conditions and average values during three time
windows. The largest profile modifications occur with the density and current
profiles, whereas the flow profiles are largely unchanged. Consistent with
experimental observations during QH-mode\cite{garofalo15}, the simulated state
leads to large particle transport relative to the thermal transport. However,
this simulation result is somewhat puzzling given the stochastic magnetic field
region within the pedestal. We posit three potential explanations that require
further investigation. The first possibility is that the simulation time is
too short for the profiles to reach a fully relaxed state. 
The second theory is that the large thermal conductivity
changes the phase of the temperature perturbation relative to the density
perturbation in such a way that the flux-surface-averaged advective transport
(where the particle flux is $(d/dV)\int_V \tilde{n} \tilde{\mathbf{v}} \cdot
\nabla \psi / |\nabla \psi| dV$ and the thermal flux is roughly 
$(d/dV)\int_V \tilde{T} \tilde{\mathbf{v}} \cdot \nabla \psi / |\nabla \psi|
dV$; here $V$ is the
volume enclosed by a flux surface) is large for the plasma density but small
for the plasma energy. A third hypothesis is that the stochastic transport is
small because the temperature in the open-field line region is large
($\simeq1\;keV$) and thus the effective temperature gradient along the
field-lines is small. For this computation, the pressure profile (and thus
implicitly the temperature profile) in the SOL was chosen to minimize the SOL
currents. Future computation will include temperatures in the open-flux region
that are at least an order of magnitude smaller. The experimentally relevant
value is somewhat difficult to determine as the ion temperature profile is not
well constrained in the open-field-line region. Given the relatively low
density outside the LCFS, the CER measurements can be corrupted by effects from
confined particles with large banana orbits from the higher-density pedestal
region.

\section{Discussion and Conclusions} 
\label{sec:conclusions}

With regard to the saturation mechanism of the unstable modes, there are two
avenues to saturation from a spectral energy perspective: (1) The unstable
perturbations can directly modify the mean fields and eliminate the source of
free energy by relaxing the profile gradients; or (2) the perturbations can
couple to stable modes that dissipate the energy (again through modification of
the mean fields or through energy flow to the boundary of the domain). 
Figure~\ref{fig:profs} shows that the perturbations in this case make
non-trivial modifications to the $n_\phi=0$ fields that relax the pressure
gradient and current profile leading to saturation. 

One complication to this picture arises when considering the steady-state
nature of the initial state from the reconstructed fields. As mentioned in
Sec.~\ref{sec:reconstruct}, these fields are assumed to be time independent
given the presence of sources, sinks and fluxes that are outside the scope of
our modeling. For a discharge state with broadband MHD activity, the
contribution of the flux from the MHD perturbations is included in the
steady-state assumption. In this sense, our modeling of the transport from the
MHD perturbations constitutes a `double counting' of this flux. For studies
that compare the level of this flux to experiment there are two approaches to
resolve this inconsistency: (1) The initial profiles could differ from the
experiment and be more unstable such that the MHD perturbations relax the
profiles to a state that resembles the reconstructed profiles; or (2) the
$n_\phi=0$ modifications from the MHD perturbations could be cancelled (or
ignored) such that the final profiles match the reconstructed values. This
first approach suffers from the difficulty of finding the more unstable state,
a priori, that relaxes to the reconstructed state.  The second approach is
the traditional way turbulent flux calculations are performed and is of interest
for future studies. As this approach eliminates mode saturation through
modification of the mean fields, the stability of the mode spectrum becomes
critical. In particular, it is likely that simulations must be performed with
an extended-MHD model that includes two-fluid, first-order
finite-Larmour-radius effects that stabilize the intermediate-$n_\phi$ modes.

Our simulations produce an MHD turbulent-like state, which is a good candidate
to at least partially explain the broadband-MHD phenomena. However, additional
comparisons to experimental data are required to confidently claim these
simulations truly model the discharge dynamics. Prior attempts to compare with
magnetic probe data proved unsuccessful as the probe measurement temporal
resolution (200 kHz) is approximately two orders of magnitude smaller
than our nonlinear simulation time period. Higher time-resolution measurements
that make local measurements of the perturbations (e.g. beam-emission
spectroscopy and Doppler reflectometry) are a more promising avenue to pursue
validation and will be the subject of future studies.


\appendix

\begin{acknowledgments}
We thank Carl Sovinec and Eric Held for discussions involving the nonlinear
dynamics of this work, Jim Myra and Dan D'Ippolito for discussions pertaining
to the SOL treatment in the initial state, Stuart Hudson and Todd Evans for
discussions regarding magnetic field structure and the reviewers for detailed
comments on the manuscript. This material is based on work supported by the
U.S.  Department of Energy Office of Science and the SciDAC Center for Extended
MHD Modeling under contract numbers DE-FC02-06ER54875, DE-FC02-08ER54972
(Tech-X collaborators) and DE-FC02-04ER54698 (General Atomics Collaborators).
This research used resources of the Argonne Leadership Computing Facility,
which is a DOE Office of Science User Facility supported under contract
No.~DE-AC02-06CH11357,  and resources of the National Energy Research
Scientific Computing Center, a DOE Office of Science User Facility supported by
the Office of Science of the U.S.  Department of Energy under contract
No.~DE-AC02-05CH11231.
\end{acknowledgments}

\bibliographystyle{apsrev4-1}
\bibliography{Biblio}

\end{document}